\documentclass[letterpaper]{jpconf}
\usepackage{color,graphicx}

\newcommand{\vect}[1]{{\bf{#1}}}
\newcommand{\f}[2]{\frac{#1}{#2}}

\newcommand{\genasis}{GenASiS}
\newcommand{\rShock}{R_{\mbox{\tiny{Sh}}}}
\newcommand{\rPNS}{R_{\mbox{\tiny PNS}}}

\begin{document}
\title{Magnetic field generation by the stationary accretion shock instability}

\author{E Endeve$^{1,2,3}$, C Y Cardall$^{1,2}$, R D Budiardja$^{2}$ and A Mezzacappa$^{1}$}

\address{$^1$Physics Division, Oak Ridge National Laboratory (ORNL), Oak Ridge, TN 37831, USA}
\address{$^2$Department of Physics and Astronomy, University of Tennessee, Knoxville, TN 37996, USA}
\address{$^3$Joint Institute for Heavy Ion Research, ORNL, Oak Ridge, TN 37831, USA}

\ead{endevee@ornl.gov}

\begin{abstract}
By adding a weak magnetic field to a spherically symmetric fluid configuration that 
caricatures a stalled shock in the post-bounce supernova environment, we explore the capacity 
of the stationary accretion shock instability (SASI) to generate magnetic fields.  
The SASI develops upon perturbation of  the initial condition, and the ensuing flow 
generates---{\em in the absence of rotation}---dynamically significant magnetic fields ($\sim 10^{15}$ G) 
on a time scale that is relevant for the explosion mechanism of core-collapse supernovae.  
We describe our model, present some recent results, and discuss their potential relevance for 
supernova models.  
\end{abstract}

\section{Introduction}

Magnetic fields in the context of core-collapse supernovae---the final evolutionary stage of
massive stars---have recently attracted considerable interest 
(see \cite{shibata_etal_2006,burrows_etal_2007}, and references therein) due, in part, 
to the generically observed asphericity of supernova explosions \cite{wang_etal_2001}.  

In particular, 
simulations of rapidly rotating progenitors followed for several hundred milliseconds past core bounce
demonstrate the ability of magnetic fields to drive and collimate bipolar explosions \cite{burrows_etal_2007}:
Differential rotation near the surface 
of the rapidly rotating proto-neutron star (PNS) winds and organizes poloidal magnetic fields 
into `magnetic towers' along the rotation axis, consisting mainly of toroidal fields.  
Although this mechanism is compelling, current models rely on 
rotation rates---and an initial field strength and topology---that
are incompatible with stellar evolution calculations 
that take magnetic effects into account \cite{heger_etal_2005}.

It is often assumed that the magneto-rotational instability (MRI), which may well operate in core-collapse supernovae \cite{akiyama_etal_2003}, will bridge the gap and make the winding mechanism work. 
Capturing the MRI in numerical simulations requires higher spatial resolution than has been possible in realistic supernova simulations. The MRI has been observed in a simplified model of stellar collapse \cite{shibata_etal_2006}, but the potential impact of the MRI (and magnetic fields generally) on the core-collapse supernova explosion mechanism remains unclear.

A different mechanism that may account for aspherical supernovae is the SASI, a
hydrodynamical instability of the stalled supernova shock wave.  Idealized two-dimensional (2D)
and three-dimensional (3D) simulations have shown the SASI to have a dramatic impact on the 
postshock flow \cite{blondin_etal_2003,blondin_mezzacappa_2007}:  The $\ell=1$ (sloshing) mode
is seen to dominate 2D simulations, while the $m=1$ (spiral) mode dominates in 3D and may
generate significant pulsar spin.  The SASI is also seen to play an important role in the dynamics of 
more realistic models \cite{buras_etal_2006,burrows_etal_2006,bruenn_etal_2006,scheck_etal_2008}.  

Given its impact on the postshock flow, it is natural to explore the influence of the SASI on 
magnetic fields (and vice-versa) in the supernova environment.  By extending 
an axisymmetric model \cite{blondin_etal_2003} 
to include magnetic fields we have initiated an investigation of 
the evolution of magnetic fields when driven by the SASI:  Starting from a spherically symmetric, 
non-rotating configuration resembling a stalled supernova shock wave, we find that 
axisymmetric
SASI-induced 
flows are able to amplify magnetic fields to dynamically significant levels within the time frame a 
supernova is expected to reach explosive conditions.  

\section{Physical model and numerical solution}

In order to study magnetic field generation by the SASI, and the potential relevance to the 
core-collapse supernova explosion, we adopt a highly idealized description of the post-bounce 
supernova environment \cite{blondin_etal_2003}:  A steady-state, spherically symmetric accretion 
shock is located at $r=\rShock=200$~km from the center of the PNS.  Matter at 
larger radii is falling into the shock at the free-fall speed, $u=\sqrt{2GM/r}$, where $G$ is Newton's 
constant, and the mass $M$ of the central object  is set to $1.2~M_{\odot}$.  A fixed accretion rate of 0.36~M$_{\odot}$~s$^{-1}$ is employed, and a constant, 
highly supersonic Mach number of 300 is used to set the pressure in the pre-shock gas.  
The gravitational potential is given by the point-mass formula  $\Phi=-GM/r$.  
The Rankine-Hugoniot conditions determine the hydrodynamic state just 
inside the shock, and the Bernoulli equation is solved for the structure from the shock to the surface 
of the PNS, which serves as the inner boundary and is located at $r=\rPNS=40$ km.  
The initial magnetic field is that of a split monopole:  $\vect{B}=B_{r} \vect{e}_{r}$, where 
$B_{r}=\mbox{sign} \left(\cos\theta\right)\times B_{0}\left(r/\rPNS\right)^{2}$, and $\theta$ is the 
polar angle in the spherical coordinate system.  The strength of the
magnetic field at $r=\rPNS$, $B_{0}$, is set to $10^{10}$~G in our base model.  The magnetic field lines are initially parallel to the 
accreting flow and the configuration is consistent with the steady state initial condition.  
The system is described by the non-relativistic ideal magnetohydrodynamic (MHD) equations, 
which are closed with a polytropic equation of state (EoS) with the ratio of specific heats $\gamma$ 
set to $4/3$, similar to collapsing matter in a supernova.  

We have implemented a time-explicit, second-order, semi-discrete central-upwind 
scheme \cite{kurganov_etal_2001} in our code, \genasis, for the integration of the time-dependent ideal 
MHD equations.  The equations are formulated in their integral form, appropriate for finite-volume shock-capturing methods.  We use the HLL formulae \cite{londrillo_delZanna_2004} 
to compute the numerical fluxes and electric fields, located on faces and edges of a computational 
cell, respectively.  Second-order spatial accuracy in smooth regions, while maintaining a 
non-oscillatory behavior near shocks and discontinuities, is achieved by linear interpolation subject to 
a generalized minmod limiter \cite{kurganov_tadmor_2000}.  The method of constrained 
transport \cite{evans_and_hawley_1988} is used to maintain the divergence-free condition on the 
magnetic field during the evolution of the MHD equations.  Second-order temporal accuracy is obtained 
with a two-step Runge-Kutta time step algorithm \cite{kurganov_tadmor_2000}.  
The equations are solved on a uniform Cartesian grid ($\Delta x=\Delta y=\Delta z \equiv\Delta l$), 
and in order to solve axially symmetric problems we have extended the `Cartoon' method \cite{alcubierre_etal_2001} to the MHD case.  

The results presented in this study have been computed with a spatial resolution given by
$\Delta l/\rPNS \simeq 0.06$.  At this resolution we obtain solutions that are consistent with 
the MHD equations in the sense that the total energy in the system is conserved at an acceptable 
level, and that the unperturbed model can indefinitely maintain (modulo a slight initial relaxation) the steady state of the initial condition, without triggering the SASI from numerical noise.  

\section{Results}

\begin{figure}[h]
  \includegraphics[width=4.2in]{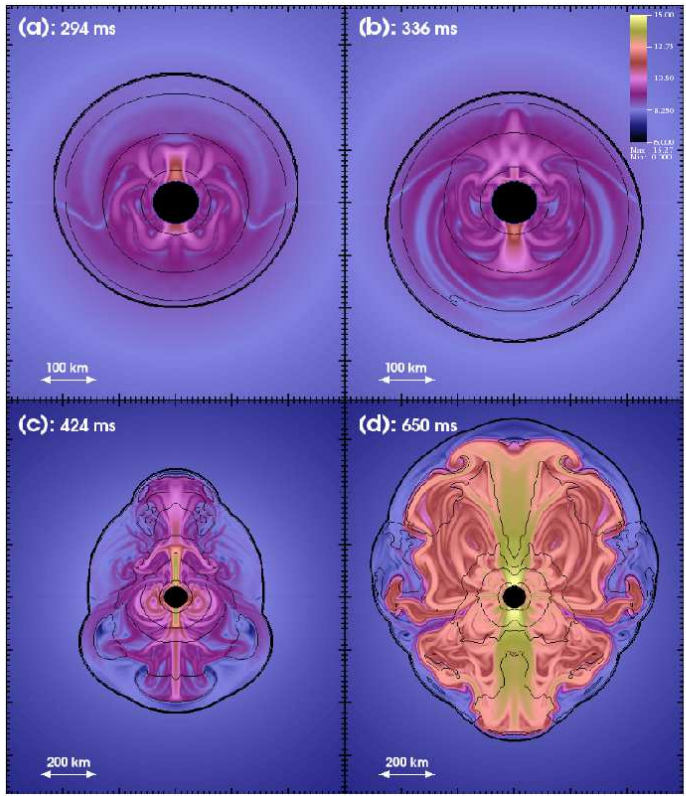}
  \hspace{-0.2in}
  \begin{minipage}[b]{2.1in}
    \caption{\label{fig:figure1} 
      Color plot of the magnitude of the magnetic field, $|\vect{B}|$, during the evolution of the SASI.  
      Panel (a) is sampled at 294~ms, panel (b) at 336~ms, while lower panels, (c) and (d), are 
      sampled at 424~ms and 650~ms, respectively.  The color-bar in the upper right 
      (valid for all panels) indicates the magnetic field strength on a log scale, extending from 
      $10^{6}$~G to $10^{15}$~G.  Thin lines are contours of constant density; starting with 
      the innermost contour we plot density contours at $\rho=10^{10}$, $10^{9}$, $3 \times 10^{8}$, 
      and $6 \times 10^{7}$ g cm$^{-3}$.  The thick contour indicates the location of the 
      shock surface.  Notice the factor of two difference in the spatial scale used in the upper 
      and lower panels of the plot.  }
  \end{minipage}
\end{figure}

In order to initiate the SASI we perturb the initial condition by placing two dense rings in the material 
ahead of the shock.  As the density enhancements pass through the shock they introduce 
non-radial perturbations that grow and initiate the SASI.  The $\ell=1$ mode quickly manifests itself 
as an up-down `sloshing' of the global accretion shock configuration.  In Figure (\ref{fig:figure1}) we 
plot a time series illustrating the evolution of the SASI from the onset of the calculation and well into the 
non-linear phase:  We show a color plot of the magnetic field strength, $|\vect{B}|$, at four different times 
during the evolution.  The snapshots shown are taken at 294~ms (a), 336~ms (b), 424~ms (c), 
and 650~ms (d).  

During SASI development the shock starts to deviate from spherical symmetry, the 
accreting flow hits the shock at an oblique angle, and significant non-radial flows are introduced 
inside the shocked cavity.  The lateral flow carries magnetic field toward the symmetry axis, where 
the strength of the field increases due to compression:  At 294~ms the strongest field is found around 
the north pole, with $B_{\mbox{\tiny{max}}}\sim 2 \times 10^{12}$~G.  At the south pole the field strength 
is about $3\times 10^{11}$~G.  Some 40~ms later, at 336~ms, the situation is reversed, and the strongest 
field, $B_{\mbox{\tiny{max}}}\sim 2\times 10^{12}$~G, is found around the south pole.  

An internal shock forms in the postshock gas around 360~ms, and is connected to the accretion shock
at a triple point \cite{blondin_mezzacappa_2007} (visible as a kink in the shock surface).  
The internal shock propagates in the north-south direction in panel (c).  Ahead of the internal shock, 
accretion-powered flow plunges toward the PNS and the axis of symmetry, causing further field growth.  
At 424~ms columns of stronger magnetic field are seen in both polar regions, with maxima of some $10^{14}$~G.  
At 650~ms extended regions of strong magnetic field have formed around both the north and south 
poles.  The strength of the magnetic field in these regions is above $10^{15}$~G near the surface of the 
PNS, and stays above $10^{14}$~G out to about 100~km beyond the surface of the PNS.  The magnetic 
field is these regions is strong enough to influence the dynamics:  From the density contours in panel (c) 
we see that low-density funnels have formed along the polar axis.  

\begin{figure}
  \begin{minipage}[b]{3in}
  \includegraphics[angle=00,height=3in,width=3in]{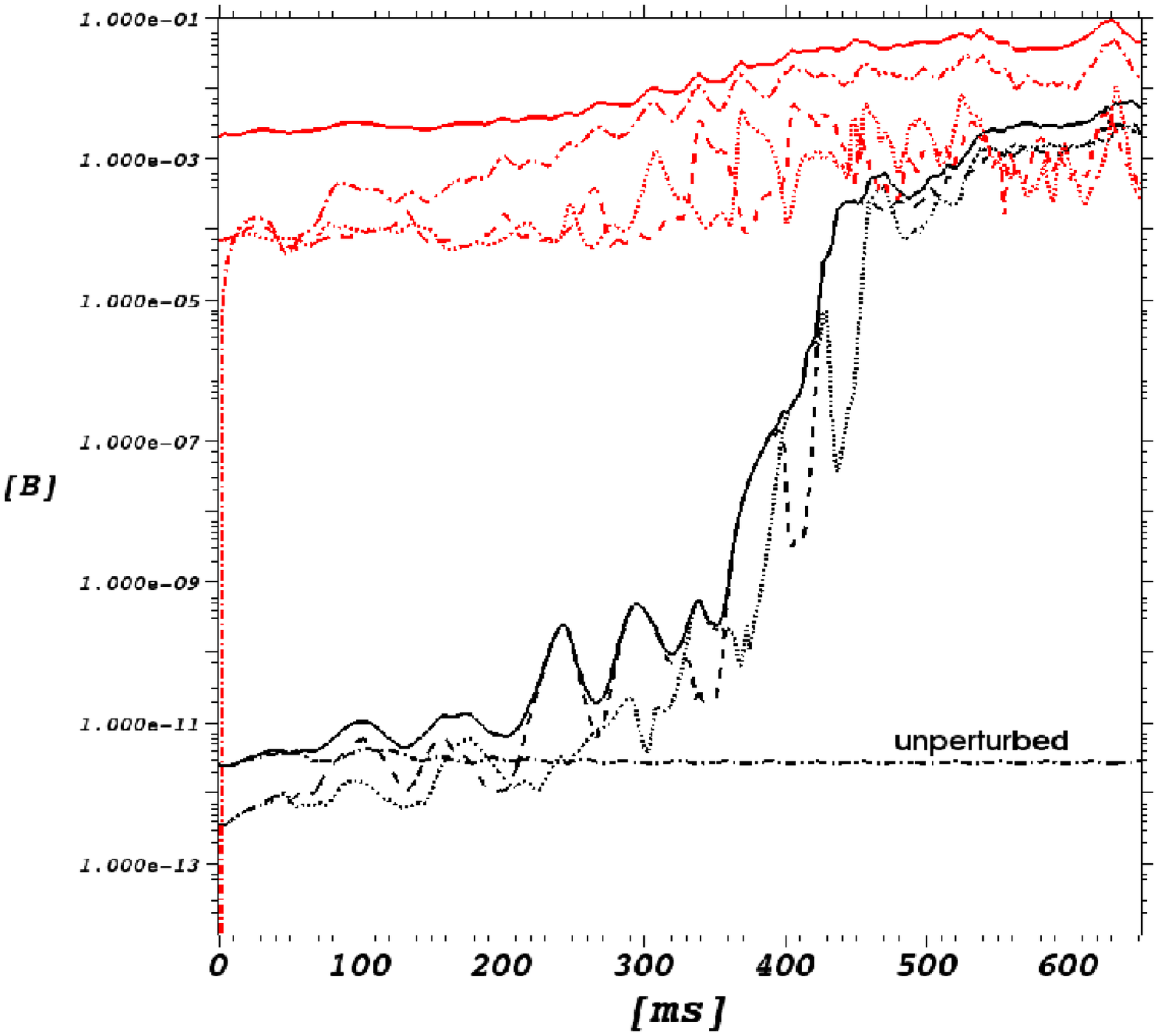}  
    \caption{\label{fig:figure2} 
      Kinetic (red curves) and magnetic (black curves) energies (see text for detail):  
      $E_{\mbox{\tiny{kin}}}^{\mbox{\tiny{Sh}}}$ (\textcolor{red}{\full}), 
      $E_{\mbox{\tiny{kin}}}^{\theta}$ (\textcolor{red}{\chain}), 
      $E_{\mbox{\tiny{kin}}}^{\mbox{\tiny{N}}}$ (\textcolor{red}{\dashed}), and 
      $E_{\mbox{\tiny{kin}}}^{\mbox{\tiny{S}}}$ (\textcolor{red}{\dotted}).  
      $E_{\mbox{\tiny{mag}}}^{\mbox{\tiny{Sh}}}$ (\full), 
      $E_{\mbox{\tiny{mag}}}^{\mbox{\tiny{N}}}$ (\dashed), and 
      $E_{\mbox{\tiny{mag}}}^{\mbox{\tiny{S}}}$ (\dotted). 
      We also plot $E_{\mbox{\tiny{mag}}}^{\mbox{\tiny{Sh}}}$ for 
      an unperturbed model (\chain).  ($1$~B=$10^{51}$~erg.)}
  \end{minipage}
  \hspace{0in}
  \begin{minipage}[b]{3in}
  \includegraphics[angle=00,height=3in,width=3in]{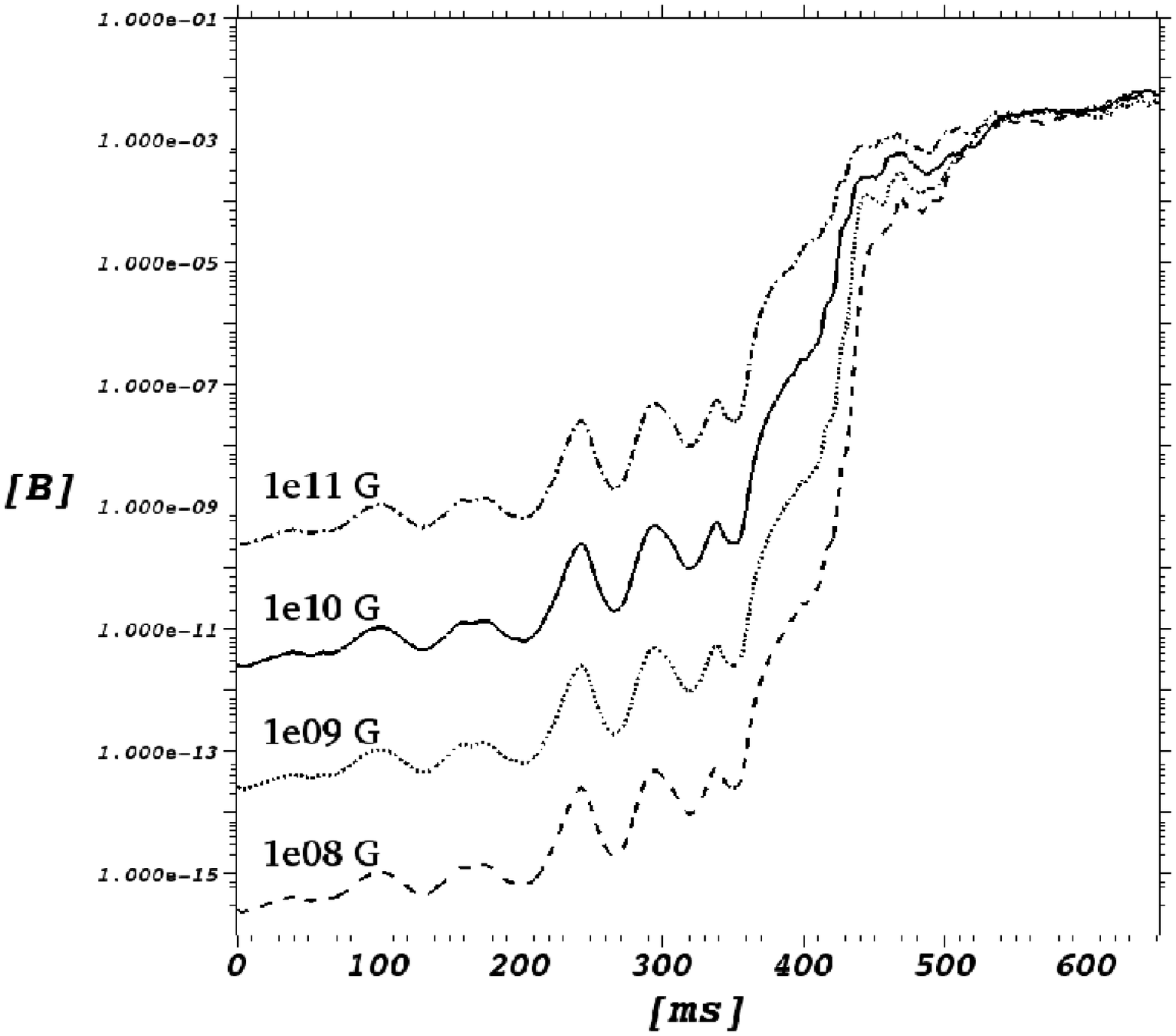}  
    \caption{\label{fig:figure3} 
      Magnetic energy inside the shocked cavity,  $E_{\mbox{\tiny{mag}}}^{\mbox{\tiny{Sh}}}$, for 
      models with various values of initial magnetic field strength $B_{0}$ at the PNS surface:  
      $10^{8}$~G (\dashed), $10^{9}$~G (\dotted), $10^{10}$~G (\full), and $10^{11}$~G (\chain), 
      respectively.  }
  \end{minipage}  
\end{figure}

The impact of the SASI-induced flow on the magnetic field is further illustrated in Figure (\ref{fig:figure2}), 
in which we plot subsets of the integrated kinetic (red lines) and magnetic (black lines)
energies inside the shocked cavity.  The solid red line represents the total kinetic energy inside the 
shock, 
$E_{\mbox{\tiny{kin}}}^{\mbox{\tiny{Sh}}}=\int_{V_{\mbox{\tiny{Sh}}}}\frac{1}{2}\rho\, \vect{u}\cdot\vect{u} \,dV$,
where 
$V_{\mbox{\tiny{Sh}}}$ is the total volume encompassed by the shock.  The red dash-dot line 
represents the kinetic energy of the lateral motion, 
$E_{\mbox{\tiny{kin}}}^{\theta}=\int_{V_{\mbox{\tiny{sh}}}} \f{1}{2}\rho u_{\theta}^{2} \, dV$.  The dashed 
and dotted red lines represent the kinetic energy around the northern and southern hemispheres,
respectively:  
$E_{\mbox{\tiny{kin}}}^{\mbox{\tiny{N}}}=\int_{V_{\mbox{\tiny{N}}}} \frac{1}{2} \rho\, \vect{u}\cdot\vect{u} \, dV$ 
and
$E_{\mbox{\tiny{kin}}}^{\mbox{\tiny{S}}}=\int_{V_{\mbox{\tiny{S}}}} \frac{1}{2} \rho\, \vect{u}\cdot\vect{u} \, dV$.  
Here $V_{\mbox{\tiny{N(S)}}}$ is the volume bounded by the surface of the PNS, the surface of the 
accretion shock, and the surface a cylinder with radius $\rPNS$, centered on the polar axis in the 
northern (southern) hemisphere.  The solid black line represents the magnetic energy inside the 
accretion shock, 
$E_{\mbox{\tiny{mag}}}^{\mbox{\tiny{Sh}}}=\int_{V_{\mbox{\tiny{sh}}}} \frac{1}{2\mu_0} \vect{B}\cdot\vect{B} \, dV$, 
while the dashed and dotted lines represent the magnetic energy in cylinders around the north 
and south pole:  
$E_{\mbox{\tiny{mag}}}^{\mbox{\tiny{N}}}=\int_{V_{\mbox{\tiny{N}}}} \frac{1}{2\mu_0} \vect{B}\cdot\vect{B} \, dV$ 
and 
$E_{\mbox{\tiny{mag}}}^{\mbox{\tiny{S}}}=\int_{V_{\mbox{\tiny{S}}}} \frac{1}{2\mu_0}\vect{B}\cdot\vect{B} \, dV$
respectively, where $\mu_0$ is the vacuum magnetic permeability.  For comparison, we also plot the magnetic energy inside the shock, 
$E_{\mbox{\tiny{mag}}}^{\mbox{\tiny{Sh}}}$, for an unperturbed model (black dash-dot line).  

The non-radial perturbations lead to the growth of the kinetic energy of the lateral motion inside the 
shock.  This is seen in the rise in $E_{\mbox{\tiny{kin}}}^{\theta}$ for $t<400$~ms.  The sloshing nature 
of the $\ell=1$ mode is also revealed in the alternating pattern of the kinetic energies contained in the 
cylinders centered around the northern and southern hemispheres.  Note the correspondence of rise 
in magnetic energies with decline in kinetic energies.  This persists to about 450~ms, when the flow 
takes on the character of the $\ell=2$ mode.  From the evolution of the magnetic energy we can see 
that the SASI has a quite dramatic impact on the magnetic field.  The early evolution is characterized 
by relatively modest oscillations in the strength of the magnetic field superimposed on a trend of 
gradual overall increase, but starting at about 360~ms, around the formation of the internal shock, 
{\em the magnetic energy increases by six orders of magnitude in roughly 100~ms}.  Beyond 450~ms the 
magnetic energy saturates at a level somewhat below the total kinetic energy inside the shock.  
The magnetic field is strongly aligned with the symmetry axis and the magnetic energy is in 
approximate local equipartition with the kinetic energy in the polar regions.  

We have also done calculations with different initial field strengths.  In Figure (\ref{fig:figure3}) we
plot the magnetic energy inside the accretion shock, $E_{\mbox{\tiny{mag}}}^{\mbox{\tiny{Sh}}}$, for 
models where we have varied $B_{0}$:  $10^{8}$~G (dashed line), $10^{9}$~G (dotted line), 
$10^{10}$~G (solid line, the model shown in Figures (\ref{fig:figure1}) and (\ref{fig:figure2})), 
and $10^{11}$~G (dash-dot line).  These calculations show that our result is insensitive to the strength 
of the initial field:  The magnetic energies saturate at the same level for all models.  

\section{Summary and discussion}

By considering an idealized model of a stalled supernova shock wave we have found that the SASI-induced 
flow beneath the shock is able to generate dynamically significant magnetic fields, in excess of $10^{15}$~G, 
on a time scale relevant to the supernova explosion, independent of the initial 
field strength.
The influence 
of the SASI on the evolution of magnetic fields in core-collapse supernovae will need to be investigated in 
more realistic models, but our idealized models offer a proof of principle that (1) magnetic fields may be amplified 
to significant strengths {\it even in the absence of rotation} and (2) compression may serve, at least in axisymmetry, as an effective 
mechanism for rapid field amplification to dynamically significant levels, contrary to conclusions 
reached in past MHD collapse calculations. Moreover, current 2D models 
\cite{buras_etal_2006,burrows_etal_2006,bruenn_etal_2006,scheck_etal_2008} demonstrate that the SASI is common 
to core-collapse supernova models.  Our results then suggest that SASI magnetic field amplification and, in 
particular, the formation of dynamically significant fields may be common as well, and not confined to special 
cases---for example, to cases of rapid progenitor rotation. The key question now is: What happens in three 
dimensions? The amplification we document here relies on the continual focusing of fluid flow toward the polar axis, 
in part an artifact of our assumed axisymmetry. Hydrodynamics simulations of the SASI in 3D yield a 
significantly different and more complex postshock flow \cite{blondin_mezzacappa_2007}, whose consequences for stellar core magnetic field amplification remain to be investigated.

\ack

This work was supported by Oak Ridge National Laboratory (ORNL), managed by UT-Battelle, LLC, 
for the DoE under contract DE-AC05-00OR22725.  This research used resources of the National 
Center for Computational Sciences at ORNL, which is supported by the Office of Science of the U.S. 
Department of Energy under Contract No. DEAC05-00OR22725.

\end{document}